# Strain-induced atomic-scale building blocks for ferromagnetism in epitaxial LaCoO$_3$

*Sangmoon Yoon[†,‡], Xiang Gao[†,‡], Jong Mok Ok[†], Zhaoliang Liao[†], Myung-Geun Han[‡], Yimei Zhu[‡], Panchapakesan Ganesh[§], Matthew F. Chisholm[§], Woo Seok Choi[¶], Ho Nyung Lee[†,*]*

[†] Materials Science and Technology Division, Oak Ridge National Laboratory, Oak Ridge, TN 37831, United States

[‡] Condensed Matter Physics and Materials Science Department, Brookhaven National Laboratory, Upton, NY 11973, United States

[§] Center for Nanophase Materials Sciences, Oak Ridge National Laboratory, Oak Ridge, TN 37831, United States

[¶] Department of Physics, Sungkyunkwan University, Suwon 16419, Republic of Korea

KEYWORDS: Epitaxial thin film, Complex oxides, Ferromagnetism, Electron microscopy

ABSTRACT

The origin of strain-induced ferromagnetism, which is robust regardless of the type and degree of strain in LaCoO$_3$ (LCO) thin films, is enigmatic despite intensive research efforts over the past decade. Here, by combining scanning transmission electron microscopy with *ab initio* density functional theory *plus U* calculations, we report that the ferromagnetism does not emerge directly from the strain itself, but rather from the creation of compressed structural units within ferroelastically formed twin-wall domains. The compressed structural units are magnetically active with the rocksalt-type high-spin/low-spin order. Our study highlights that the ferroelastic nature of ferromagnetic structural units is important for understanding the intriguing ferromagnetic properties in LCO thin films.

Epitaxial strain is one of the most effective means to manipulate the electronic and magnetic ground states of transition metal oxides (TMOs), via strong coupling between charge, spin, orbital, and lattice degrees of freedom.[1] Lattice mismatch on the order of a few percent between an epitaxially strained film and a substrate can produce several GPa of biaxial pressure. Specifically, the 1.8% compression of in-plane lattice parameters of bulk $LaCoO_3$ (LCO) requires an external pressure of 10 GPa.[2-4] Epitaxial strain is an effective means to discover new behaviors or phases, to improve materials performances or properties, and to fine-tune various functionalities.[5-7] Structural modification or crystallographic symmetry change are major reactions of matter to epitaxial strain.[8, 9] Such changes manifest in ferroelastic materials, yielding intriguing modifications of crystal structures or orientations, i.e. twin domains.[10, 11] LCO is a prototypical ferroelastic perovskite oxide,[12-14] which exhibits a monoclinic twin domain structure in bulk, resulting in twin phases with spontaneous strain.[15, 16] Moreover, since $Co^{3+}$ ions with $3d^6$ electrons can possess non-zero local spin configurations, it is an interesting model system for the study of the coupling between the lattice structure and magnetic properties. While bulk LCO is non-magnetic, epitaxially strained LCO thin films exhibit robust ferromagnetism with a phase transition temperature ($T_c$) of 80 K,[17-19] just above the liquid nitrogen temperature. In addition, whereas conventional ferromagnetism arises in metallic states, LCO is a insulator with a band gap of 0.7 eV.[20-22] The insulating behavior suppresses the charge degree of freedom, which simplifies the identification of the spin-lattice coupling.

Despite extensive research efforts over the past decade, the origin of ferromagnetism in strained LCO thin films remains a subject of debate.[19, 23-26] Whereas several scenarios were reported to account for the mechanism governing the onset of the magnetism, a consensus has not yet been achieved. Among major experimental observations, one of the most distinct features in epitaxial LCO films is a dark-striped pattern observed when imaging with high-angle annular dark field scanning transmission electron microscopy (HAADF STEM).[19, 23, 27-31] Two distinct models were developed to explain the origin and nature of such stripe patterns and their possible connection to ferromagnetism. The first model is oxygen vacancy ordering.[23, 30, 32] The dark stripes are suggested as oxygen-deficient plane inducing a complex electronic structure reconstruction with ferromagnetic order. The second model is the intrinsic ferroelastic nature of LCO.[19, 27, 28] In this case, the dark stripes were considered as stoichiometric twin walls strongly coupled to the ferromagnetism.[19, 28, 29] The details of the dark stripes have been considered as the key to the debate, but the delicate nature of the dark stripes under an electron beam makes decisive electron energy loss spectroscopy (EELS) experiments challenging; high dose electrons cause the dark stripes to be further reduced (see Fig. S1 in the Supporting Information).[27, 28] Spin-orbital superstructure[24, 26, 33] and symmetry mismatch[25] have also been suggested to account for the ferromagnetism. These models commonly considered epitaxial LCOs as ideally stretched homogeneous thin films, supported by HAADF STEM images without any dark stripes.[25, 29] However, it should be noted that the dark stripes would not be observed in HAADF STEM images if the beam direction is not parallel to the dark stripe planes (see Fig. 2S in the Supporting Information).[27] Recently, the internal structure suggested by the dark-striped HAADF STEM images was further verified by polarized neutron reflectometry.[34] Therefore, the ferromagnetism of epitaxial LCOs is believed to be explained within the dark-striped atomic structure.

The twin-wall model is supported systematically by several experimental probes over the oxygen-vacancy-order model[19, 29, 35, 36], but the mechanism to support the long-range interaction between the twin walls still has to be developed. The developed mechanism needs to account for all aspects of the unusual magnetic properties emerging from epitaxially strained LCO thin films.

One of the most intriguing properties is that $T_c$ remains almost fixed at ~80 K, regardless of the degree of strain imposed by lattice mismatch.[18, 19, 30, 32, 33] This is highly unusual considering the strong spin-lattice coupling in TMOs in general, as the strength of ferromagnetic exchange interaction should strongly depend on the orbital overlap determined by the atomic spacing. Another intriguing property is that the magnetic easy-axes are fixed to the in-plane direction regardless of the direction of epitaxial strain[14, 18, 37] This phenomenon also suggests that the ferromagnetism is governed by a specific atomistic origin, not by the global strain, in epitaxially strained LCOs.

In this paper, we further study nanoscale twin structures formed in a ferroelastic LCO thin film grown on a STO substrate by HAADF STEM for an accurate understanding of the source of ferromagnetism in otherwise nonmagnetic material when unstrained. In combination with DFT+$U$ calculations, we report that the ferroelastic structural units under significantly large local compression sandwiched by the dark stripes are responsible for the strain-induced ferromagnetism in strained LCO thin films, rather than the dark stripes themselves.

Fig. 1(a) shows a high-resolution HAADF STEM image of a LCO thin film grown on a STO substrate with a STO capping layer. The capping layer was employed both to prevent any surface degradation[36] and to check the interfacial connectivity of the distinctly visible dark stripes in HAADF images of strained films. Unconventional strain relaxation occurring in the tensile-strained ferroelastic LCO thin films manifests the formation of various structural units.[19] Near the interface, a strained region without dark stripes was observed. Above this region, a structurally modulated film's interior with distinct vertical dark stripes was clearly visible. Note that the striped region maintained the same average in-plane lattice constant as that of the STO substrate, even though the thickness is much larger than the conventional critical thickness of strain relaxation, which is on the order of a few nanometer.[38-40] Interestingly, an interfacial LCO layer without dark stripes was also observed just beneath the STO capping layer with a comparable thickness as that formed at the interface with the STO substrate. This result indicates that the non-striped interfacial region is always formed as a transition layer to connect the two materials with different symmetries (the analysis of transition interface layers based on the La-La distances will be discussed below). For the LCO/STO system, the interface layer is two to three u.c. in thickness.[19]

For more detailed structural analyses, we measured the distance of La (or Sr) columns and mapped the distance along the in-plane [100] and out-of-plane [001] directions in Fig. 1(c) and (d), respectively. These results are presented schematically in Fig. 2. The detailed procedure to measure La-La (or Sr-Sr) distances in HADDF STEM images is described in the Supporting Information. The reliability of HAADF STEM image analyses was checked by measuring the Sr-Sr distance of the STO substrate, which were 3.90 ± 0.06 and 3.91 ± 0.05 Å for in-plane and out-of-plane directions (see Fig. S3 in the Supporting Information). For reference, the pseudocubic lattice parameter of bulk LCO is 3.81 Å. The red line in Fig. 1(b) shows the distribution of La-La distances along the in-plane direction, revealing three distinct structural units. The distances of 3.68, 3.82, and 4.46 Å correspond to three different structural units within the film's interior. The large La-La distance of 4.46 Å obviously represents the structural unit for dark stripes, i.e. twin walls. Importantly, the distances of 3.68 and 3.82 Å both appear at the bright region within the film's interior, which will be discussed in more details in the following paragraph (compare Fig. 1(c) to Fig. 1(a)). In this paper, we will denote the units with distances of 3.68, 3.82, and 4.46 Å as compressed-unit *(c*-unit*)*, bulk-like-unit *(b*-unit*)*, and tensile-stretched-unit *(t*-unit*)*, respectively. Herein, the LCO structural unit with the lattice parameter of STO, 3.91 Å, is not clearly observed in Fig. 1(b). Fig. 1(c) shows that the interface layer is composed of diluted ferroelastic units, not directly tetragonal with the in-plane lattice parameter of 3.91 Å. In other words, the interface layer is formed as a transition layer between the ferroelastic LCO and the non-ferroelastic cubic STO. Note that the transition interface layer is formed not only above the substrate but also beneath the capping layer; the thickness of capping layers in our samples is about 10 u.c.. The green line in Fig. 1(b) shows the distribution of La-La spacing along the out-of-plane direction. The La-La distances of the $c$-, $b$-, and $t$-units are universally 3.86 Å, while those of the interface unit ($i$-unit) is 3.75 Å, indicating the ferroelastic transition layers have a similar nature to the conventional interface layers in terms of the contraction along the out-of-plane direction.[22] Interestingly, the in-plane lattice mapping shown in Fig. 1(c) also shows that the structural modulation of the ferroelastic LCO is propagated into both STO substrate and capping layer over a few u.c..

Owing to the ferroelastic nature of LCO, the phase within the film interior was found to be further separated into two different nano-scale domains, i.e., a twin-wall superlattice phase and a bulk-like phase. Specifically, the La-La distance map of the in-plane direction (Fig. 1(c)) showed that the bright region in the HAADF STEM image should be separated into two different kinds of structural units, i.e. $c$-unit and $b$-unit. The $c$-unit always appears next to the $t$-unit with a thickness of two u.c., forming a superlattice structure with a three-u.c. periodicity together with the dark stripes. Note that the superlattice is always terminated by the $c$-unit, not by the $t$-unit. It should be emphasized that the LCO superlattice with three u.c. periodicity has been frequently observed not only in strained epitaxial thin films, but also in bulk samples, such as ball-milled stressed micropowders.[15, 16] Based on electron diffraction experiments, Vullum *et al.* pointed out that the spontaneous superlattice structure under external pressure results from the periodic occurrence of twin domains.[16] The coherent observation both in thin films and bulk samples strongly supports that this superstructure is a result of an intrinsic structural response of the ferroelastic LCO. Following their interpretation, we denoted this periodic superlattice as a twin-wall superlattice. We note again that the superstructures are absent in some STEM images of LCO thin films, possibly due to their periodic orientation along the [010] direction (see Fig. S2 in the Supporting Information).[27] The *b*-unit appears in the bright regions wider than the width of three u.c. This *b*-unit forms an independent phase with nano-scale domain, clearly distinguished from the twin-wall superlattice with the dark stripes. Since the latter unit has La-La distances similar to the lattice parameter of bulk LCO, we denoted the corresponding phase as the bulk-like phase. The inhomogeneous occurrence of dark stripes in HAADF STEM images[19, 37] implies the coexistence of the twin-wall superlattice and bulk-like phases within the film interior.

Having identified the individual structural units of LCO thin film using STEM, we employed DFT + $U$ calculations to characterize the magnetic ground state.[26] In particular, the total energies of the four plausible magnetic configurations, i.e. G-type high spin (HS) - antiferromagnetic (AFM), rocksalt-type HS/low spin (LS)- ferromagnetic (FM), HS-FM, and LS nonmagnetic (NM) states (see the schematics in Table I), were compared for the bulk phase composed of observed structural units. As summarized in Table 1, the bulk phase composed of *c*-, *b*-, *t*-, and *i*-units are considered. Here, the *i*-unit is simplified to a tetragonal unit with the in-plane and out-of-plane lattice parameter of 3.91 and 3.75Å. For the *b*-, *t*-, and *i*-units, the G-type HS-AFM order was the most stable magnetic configuration. On the other hand, the rocksalt-type HS/LS-FM order was the most stable state for the *c*-unit. Indeed, only the bulk phase consisting of the compressed units (*c*-units) within the twin-wall superlattice was magnetically active, among the considered structural phases. Note again that the rocksalt-type HS/LS-FM order was recently suggested based on resonant X-ray scattering measurements, where the superexchange between HS and LS Co$^{3+}$ was responsible for the ferromagnetic interaction of LCO thin films.[26] Consistently, the HS/LS-FM order is much more stable than the other magnetic configurations within the *c*-units; 53 and 348 meV/f.u. lower than the HS-AFM and the LS-NM state, respectively. In addition, the numerical result implies that the ferromagnetic properties of the *c*-units will disappear if these units are diluted into the tetragonal-like structures at the interface. The density of states (DOS) of the HS/LS-FM and HS-AFM orders yield insulating characters with a band gap of about 1.0 eV, consistent with the experiment (See Fig. S4 in Supplementary Information). On the contrary, the DOS of the HS-FM order and the LS-NM state exhibited a metallic character for all the bulk phases.

To further assess the feasibility of the long-range interaction of rocksalt-type HS/LS-FM order in the realistic structure of a strained LCO film, the magnetic ground state of the twin-wall superlattice was studied. In particular, we considered the three plausible magnetic configurations for the spontaneous twin-wall superlattice with three u.c. periodicity, i.e. HS/LS-FM model, HS-AFM model, and HS/LS/LS (HLL) FM model, as shown in Table 2. The HS/LS-FM and HS-AFM models are the magnetic configurations where the HS-AFM and HS/LS-FM orders are stabilized in the *c*-units of two u.c. thickness, whereas the HS-AFM configuration is commonly set for the *t*-units. The HLL model is the magnetic configuration that the HS-FM and LS-NM states are stabilized in the *t*- and *c*-units, respectively. The total energies of the three different magnetic configurations are summarized in Table 2. Upon comparison, the most stable magnetic configuration was the HS/LS-FM model. The HLL model is believed to be much more unstable than the HS/LS-FM model because the LS-NM state is difficult to be stabilized in the *c*-units as shown in Table 1. This result indicates that the rocksalt-type HS/LS-FM order is likely to be stabilized in the *c*-units in the STEM-observation-compatible twin-wall superlattice.

Experimental evidence of the magnetically active twin-wall superlattices were further sought by a quantitative comparison of the magnetic properties while varying the thickness of LCO layers. We designed [STO$_{15}$/LCO$_n$]$_m$ superlattices with various LCO thicknesses ($n$) and superlattice repetitions ($m$). The total thickness of the superlattices was maintained constant. According to the HAADF STEM image analyses, the contribution of interfacial transition layer (including the upper and lower interfaces) was six u.c. per a LCO layer sandwiched by STO layers, and the rest of the thickness corresponded to the film's interior. To remove possible exchange interaction between the LCO layers, the thickness of the STO layers was fixed to 15 u.c. As shown in Fig. 3, the $n = 15$ and 8 superlattices clearly showed a ferromagnetic behavior with identical $T_c$ and coercive field ($H_c$), i.e. $T_c \approx 80$ K and $\mu_0 H_c \approx 3.0$ kOe, while $n = 5$ and 3 superlattices did not show any discernible magnetic transition and hysteresis loop. This result demonstrates that the critical thickness to activate the ferromagnetism in epitaxial LCOs is associated with the formation of the ferroelastic film's interior. Furthermore, the HAADF STEM image analyses showed that the twin-wall superlattice and the bulk-like phase occupied 78% and 22% of the film's interior in epitaxial LCOs grown on STO substrates (see Fig. S5 in Supplementary Information). In addition, the magnetic moment of HS/LS-FM model suggested by DFT+$U$ calculations was 1.33 $\mu_B$/Co, i.e., two 4$\mu_B$ ferromagnetic components per six Co sites. Based on these volume fraction and magnetic moment of the twin-wall superlattices, we estimated the saturation magnetic moment ($M_s$) of $n = 15$ and 8 superlattices. The estimated $M_s$ of $n = 15$ and 8 superlattices were 0.63 and 0.27 $\mu_B$/Co, respectively, which are in good agreement with the experimental values. (see Fig. 3(c) and Fig. S6 in Supplementary Information). This result indirectly verifies that the twin-wall superlattices within the film's interior are magnetically active regions in epitaxially strained LCOs.

The results obtained through STEM image analyses, DFT + $U$ calculations, and LCO/STO superlattice experiments provide direct insights into the ferromagnetic properties of epitaxially strained ferroelastic LCO thin films. First, the $c$-unit in the twin-wall superlattice is the magnetically active region rather than the $t$-units, i.e. dark stripes, previously thought to be. The long-range interaction between the dark stripes (across the three u.c. width) is not ferromagnetic in this case. Rather, the short-ranged magnetic interaction within the nano-scale domain prevails, originating from the spontaneous periodic twin-wall structure. Within this model, the unusual behaviors of $T_c$ and easy-axis could also be explained. Since the twin-wall superlattice is created due to the ferroelastic response to the bi-axial strain, the volume and shape of $c$- and $t$-units that compose the twin wall structure will be the same, independent of the degree of epitaxial strain or the type of substrate. The strength of exchange interaction, and hence the $T_c$ will be identical. Only the magnetization value can be modified depending on the concentration of twin-wall superlattice domains.[18, 19] It further implies that to change or enhance $T_c$, one needs to modify the ferroelastic structural unit, which is rather robust against stress. Previously, it was shown that the twin walls were propagated in a different direction in tensile- and compressive-strained LCO films, forming vertical and horizontal twin-wall superlattices, respectively.[19, 30, 32, 37] Importantly, in both strain cases, we found that the elongated bonds in $c$-units were always aligned along the in-plane direction (see Fig. S7 in Supplementary Information).[37] This observation may explain the universal in-plane magnetic easy-axis in both compressive- and tensile-strained LCO films with the stripe domains aligned horizontally and vertically, respectively.

In summary, we identified several ferroelastic structural units in the epitaxially strained LCO thin films, which was indispensable for the accurate understanding of the source of ferromagnetism. Microstructural analyses of the ferroelastic structural units and magnetic structure determined by HAADF STEM and DFT+$U$ calculations have provided important evidences to account for ferromagnetism; highly compressed units ($c$-unit) within the twin-wall superlattice domain are responsible for the strain-induced ferromagnetism in LCO thin films with HS/LS-FM order. The magnetically active $c$-units are created due to the ferroelastic response to the bi-axial strain with nano-scale phase separation, indicating that the volume and shape of distinctive structural units will be robust once the ferroelastic deformation occurred in LCO films. These results explain the unusual ferromagnetic properties of strained LCO thin films, i.e. the fixed exchange interaction yielding persistent $T_c$, the universal in-plane easy-axis, and the correlation between domain walls and magnetization. Our study highlights the dynamic role of strain in the emergence of novel electronic/magnetic state in correlated metal oxides.

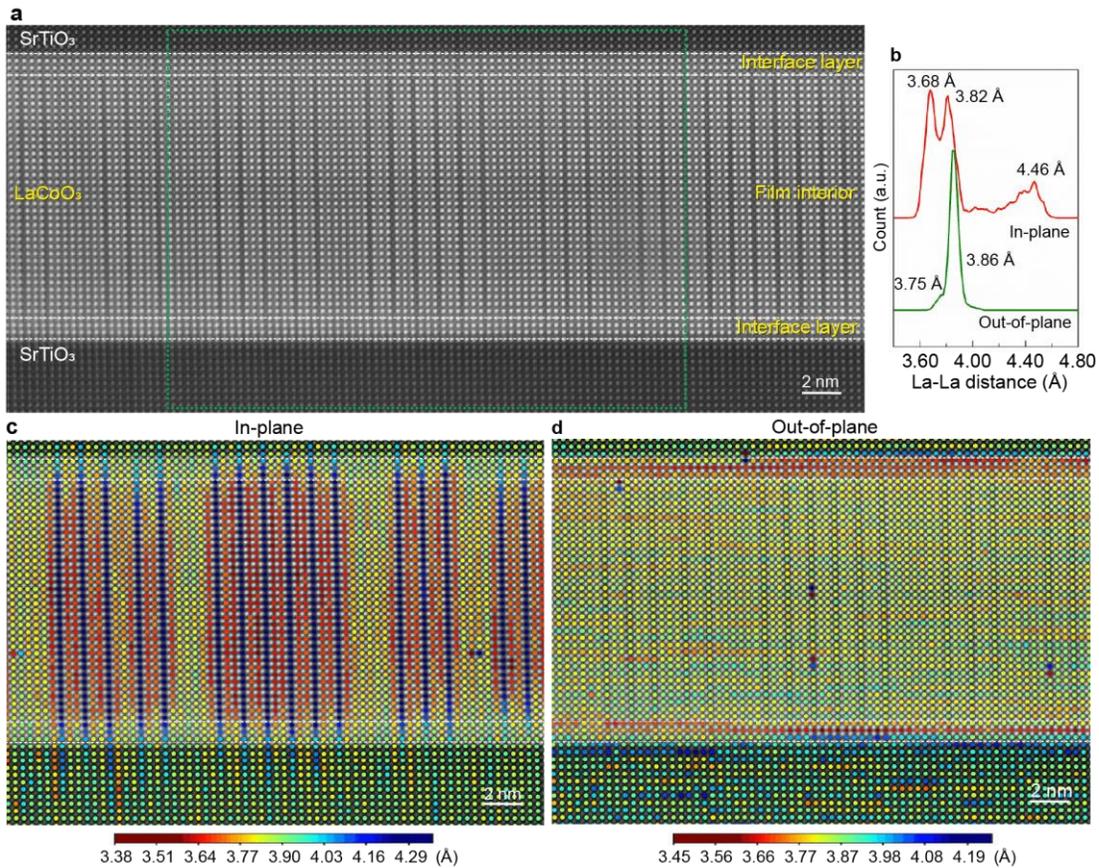

**Figure 1.** Distinctive structural units in a tensile-strained LCO epitaxial film. (a) High-resolution HAADF STEM image of a LCO thin film grown on a STO substrate with a STO capping layer on top. (b) The distribution of the La-La distances of the LCO thin film in the in-plane and out-of-plane directions. The La-La (or Sr-Sr) distance map projected in (c) in-plane [100] and (d) out-of-plane [001] directions. The green dotted box in (a) indicates the region where the La-La distances were analyzed in (c) and (d).

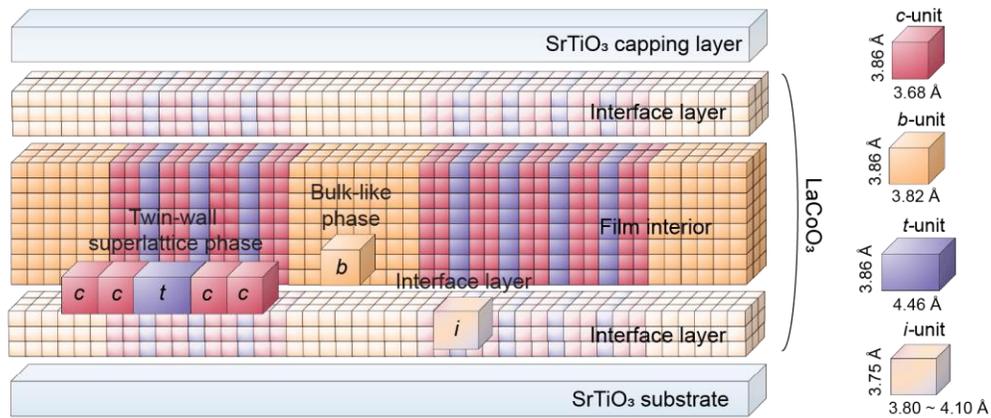

**Figure 2.** Schematic illustration of the complex internal structure of a tensile-strained LCO thin film with the identification of each structural unit. The LCO film consists of three different kinds of regions, i.e. the twin-wall superlattice composed of the compressed-unit *(c*-unit*)* and tensile-stretched-unit *(t*-unit*)*, the bulk-like phase composed of the bulk-like-unit *(b*-unit*)*, and the interface layer with diluted ferroelastic structural units (interface units *(i*-units*)*). The *c*-, *b*-, *t*-units are denoted by red, orange and blue colors, respectively, and the *i*-unit is denoted by lighter colors, indicating diluted ferroelastic structures. The lattice parameters of each structural region obtained from the HAADF STEM image analyses are also provided based on our STEM observations.

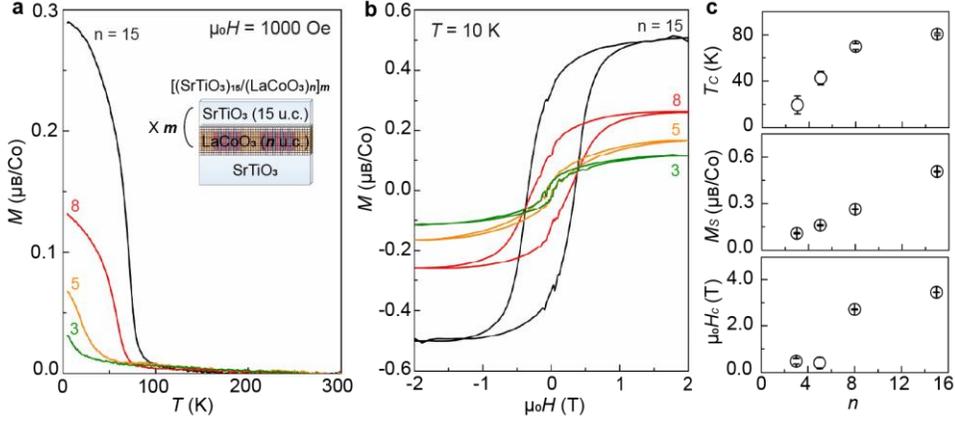

**Figure 3.** Magnetic properties of STO/LCO superlattices with different thickness of the LCO layer. (a) $M(T)$ and (b) $M(H)$ curves for $[STO_{15}/LCO_{15}]_1$ (black), $[STO_{15}/LCO_8]_2$ (red), $[STO_{15}/LCO_5]_3$ (orange), and $[STO_{15}/LCO_3]_5$ (green). (c) Transition temperature ($T_c$), saturated magnetization ($M_s$), and coercive field ($H_c$) of STO/LCO superlattices at different thickness of LCO. The proportion of the interfacial transition layer in the entire film increases as the thickness of LCO layers in superlattices decreases. $M(T)$ is measured in an applied field $H = 1000$ Oe, and $M(H)$ is measured at temperature of 10 K.

|  |  | HS-AFM | HS/LS-FM | HS-FM | LS-NM |
|---|---|---|---|---|---|
| Energy (meV/f.u.) | c-unit | 0 | -53 | 271 | 295 |
|  | b-unit | 0 | 14 | 299 | 32 |
|  | t-unit | 0 | 203 | 142 | 1094 |
|  | i-unit | 0 | 6 | 308 | 365 |

**Table 1.** Total energies of the ferroelastic structural units in four different magnetic states. Each magnetic state is schematically illustrated in the first row of the Table; Co ions and magnetic configurations are only illustrated in the schematics. Navy and grey spheres in the schematics indicate high-spin (HS) and low-spin (LS) $Co^{3+}$ ions, respectively. The energy of each magnetic state is represented relative to that of the HS AFM configuration, which is set to zero.

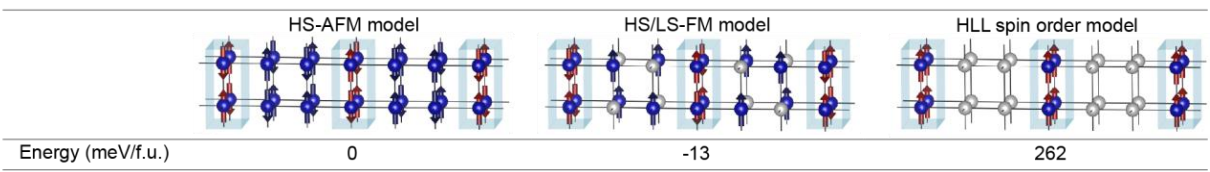

**Table 2.** Total energies of the plausible magnetic models for the twin-wall superlattice phase. Each magnetic model is schematically illustrated in the first row of the Table; Co ions and magnetic configurations are only illustrated in the schematics. Detailed description of magnetic models are given in the text. The *t*-units (dark stripes) are denoted by the transparent blue rectangular parallelepiped in each magnetic model. The energy of each magnetic model is represented relative to that of the HS AFM model, which is set to zero.

ASSOCIATED CONTENT

**Supporting Information**.

The Supporting Information is available free of charge via the internet at http://pubs.acs.org.

Experimental/computational methods; Co–$L_{2,3}$ edge EELS spectra with different electron dose condition; HAADF STEM images of a LCO thin film viewed along different directions; Distribution of Sr-Sr distances of the STO substrate; Density of states of LCO in various magnetic states; Volume fraction of twin-wall superlattice and bulk-like phase in an epitaxially strained LCO grown on a STO substrate; Volume fraction of twin-wall superlattice, bulk-like phase, and interfacial layers in n = 15 and 8 superlattices; Lattice parameters of the c-units within vertical and horizontal twin-wall superlattice and magnetic easy axis of epitaxial strained LCOs

AUTHOR INFORMATION

**Corresponding Author**

* Corresponding author: Ho Nyung Lee, Email:  hnlee@ornl.gov

**Author Contributions**

‡ These authors contributed equally. S.Y., X.G., and H.N.L designed the study. S.Y., X.G. and M.G.H. carried out and analyzed STEM experiments. S.Y. and P.G. conducted DFT calculations. J.M.O. and Z.L. grew the films and measured magnetic properties. S.Y., W.S.C., and H.N.L. wrote the manuscript. All authors discussed and commented critically on the manuscript. All authors have given approval to the final version of the manuscript.

ACKNOWLEDGMENT

This work was supported by the U.S. Department of Energy, Office of Science, Basic Energy Sciences, Materials Sciences and Engineering Division. P.G. for part of theory was supported by the Center for Nanophase Materials Sciences, which is a DOE Office of Science User Facility. W.S.C. was supported by the Basic Science Research Programs through the National Research Foundation of Korea (NRF) (NRF-2019R1A2B5B02004546).

ABBREVIATIONS

STEM, scanning transmission electron microscopy; DFT+U, density functional theory *plus U*; FM, ferromagnetism; AFM, antiferromagnetism; NM, nonmagnetism; HS, high-spin; LS, low-spin

# Strain-induced atomic-scale building blocks for ferromagnetism in epitaxial LaCoO$_3$

*Sangmoon Yoon[†,‡], Xiang Gao[†,‡], Jong Mok Ok[†], Zhaoliang Liao[†], Myung-Geun Han[‡], Yimei Zhu[‡], Panchapakesan Ganesh[§], Matthew F. Chisholm[§], Woo Seok Choi[¶], Ho Nyung Lee[†,*]*

[†] Materials Science and Technology Division, Oak Ridge National Laboratory, Oak Ridge, TN 37831, United States

[‡] Condensed Matter Physics and Materials Science Department, Brookhaven National Laboratory, Upton, NY 11973, United States

[§] Center for Nanophase Materials Sciences, Oak Ridge National Laboratory, Oak Ridge, TN 37831, United States

[¶] Department of Physics, Sungkyunkwan University, Suwon 16419, Republic of Korea

EXPERIMENTAL/COMPUTATIONAL METHODS

High-quality LCO thin films and superlattices were epitaxially grown on the STO substrate using pulsed-laser deposition. Samples were grown at 700 °C in oxygen partial pressure of 100 mTorr. Cross-sectional TEM specimens were prepared using ion milling after conventional mechanical polishing. HAADF STEM measurements were performed on Nion UltraSTEM200 operated at 200kV. The microscope is equipped with a cold field emission gun and a corrector of third- and fifth-order aberration for sub-angstrom resolution. The collection inner half-angle for HAADF STEM were 65 mrad. Noise arising in the high-resolution STEM images was reduced with an adaptive Wiener filter.[38] The La-La (or Sr-Sr) distances in HAADF STEM images were measured as follows: First, A-site atomic columns were detected in a HAADF STEM image using the blob-detection function in skimage library. Second, coordinates of each atomic column were refined by the fitting of two dimensional gaussian functions. Third, distances of neighboring A-site columns were estimated with the measured atomic coordinates. Finally, La-La (or Sr-Sr) distances were averaged for each unit and the averaged value was represented on a B-site atom (Co or Ti) with color code. All image analysis was performed using python libraries (numpy, scipy, and skimage).

*Ab initio* DFT calculations were performed using the Vienna ab initio simulation package (VASP) code.[1] The Perdew–Burke–Ernzerhof plus Hubbard correction (PBE+$U$+$J$) was used for the exchange-correlation functional,[2] in which the double-counting interactions were corrected using the full localized limit (FLL).[3] The Hubbard correction was employed in this work to correct the spurious self-interaction error which causes the localized $d$ orbital to be improperly delocalized.[4] The values used for the on-site direct Coulomb parameter ($U$) and the anisotropic Coulomb parameter ($J$) were 4.5 and 1.0 eV, respectively.[5] A plane wave basis set at a cutoff energy of 600 eV was used to expand the electronic wave functions, and the valence electrons were described using the projector-augmented wave potentials. All atoms were relaxed by the conjugate gradient algorithms until none of the remaining Hellmann–Feynman forces acting on any atoms exceeded 0.02 eV Å$^{-1}$. The magnetization was determined with a 7T Quantum Design MPMS3 measured using conventional techniques from hysteresis loop measurements after subtracting a linear background to correct for the diamagnetic response of the STO substrate. $T_c$ of [LCO$_n$/STO$_{15}$]$_m$ superlattices is determined by the intersection point of the two tangents around the inflection point of $M(T)$ curves.

SUPPLEMENTARY FIGURES

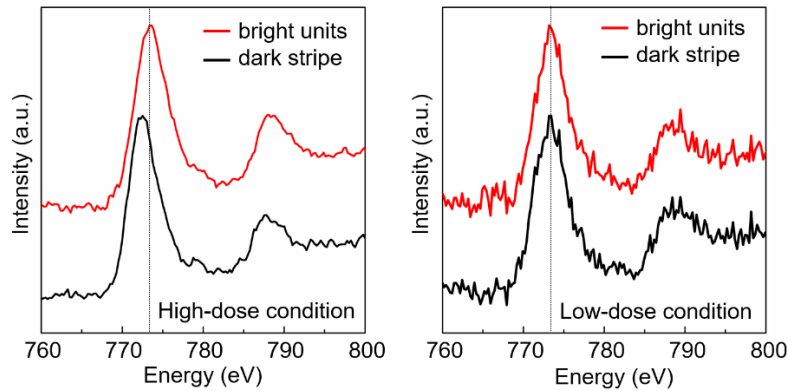

**Figure S1.** Co–$L_{2,3}$ edge electron energy loss (EEL) spectra of dark stripes and bright units with different electron dose condition; (a) high- and (b) low-dose conditions. As a high-dose condition, the EELS spectrum was obtained by exposing for a long time at a specific position (25 sec/1 point). A low-dose EELS spectrum was achieved by integrating short-exposure-time spectra obtained at equivalent but various points (0.1 sec/1 point × 100 points). The high-dose EELS spectrum has advantages in terms of signal-to-noise ratio but can be easily damaged by the injected electron beam. Interestingly, in the high-dose condition, the Co–$L_3$ edge of a dark stripe is red-shifted from that of a bright unit (here, *b-unit*). Meanwhile, in the low-dose condition, the positions of Co–$L_3$ edge are identical in all regions including dark stripes and bright units. These results demonstrate that delicate nature of dark stripes; The EEL spectra of dark stripes are particularly sensitive to the electron-dose conditions. In addition, it verifies that the pristine dark stripes are not associated with oxygen vacancies.

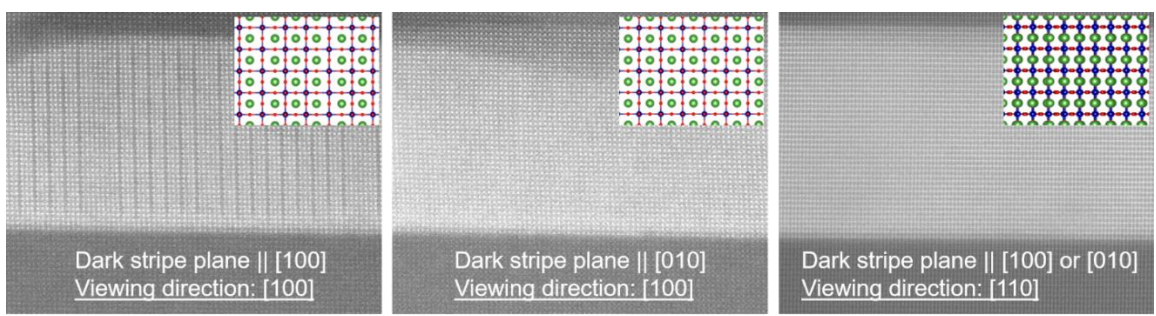

**Figure S2.** High-resolution HAADF STEM images of a LCO thin film viewed along different directions; (a)-(b) the [100] and (c) [110] zone axis. The STEM specimens with the different zone axis were prepared using the same LCO thin film. The dark stripes are seen only when the dark stripe and viewing direction are parallelly aligned. If the dark stripes are aligned in the [010] direction while the viewing direction is the [100] direction, the dark stripes cannot be seen. In addition, the dark stripes are not seen when the film is viewed along the [110] direction. It indicates that the dark stripes cannot be seen in some HAADF STEM images with unsuitable zone axis. Furthermore, these images suggest that the dark stripes always exist in the film.

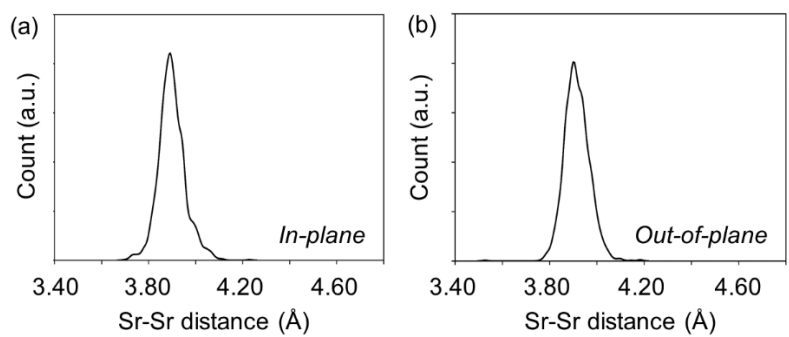

**Figure S3.** The distribution of Sr-Sr distances of the STO substrate in (a) the [100] direction and (b) the [001] direction. The average value and standard variation of in-plane and out-of-plane Sr-Sr distances are 3.90 ± 0.06 and 3.91 ± 0.05 Å, respectively.

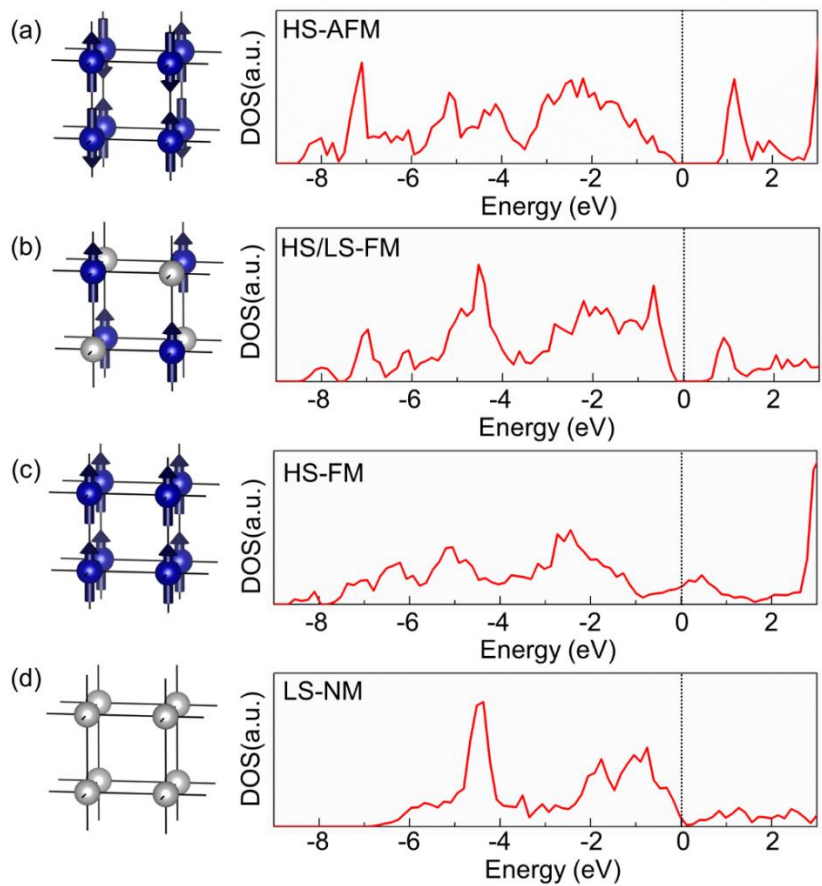

**Figure S4.** Density of states (DOS) of LCO in various magnetic states. DOS of LCO in (a) HS-AFM, (b) HS/LS-FM, (c) HS-FM, and (d) LS-NM. HS AFM and HS/LS FM LCO exhibit insulating behavior, whereas HS FM and LS NM LCO shows metallic behavior. The above DOSs are the results computed for the bulk phase composed of the compressed unit (*c-unit*) in twin-wall superlattice. The DOS computed in the other bulk phases also show a similar trend with the above DOSs.

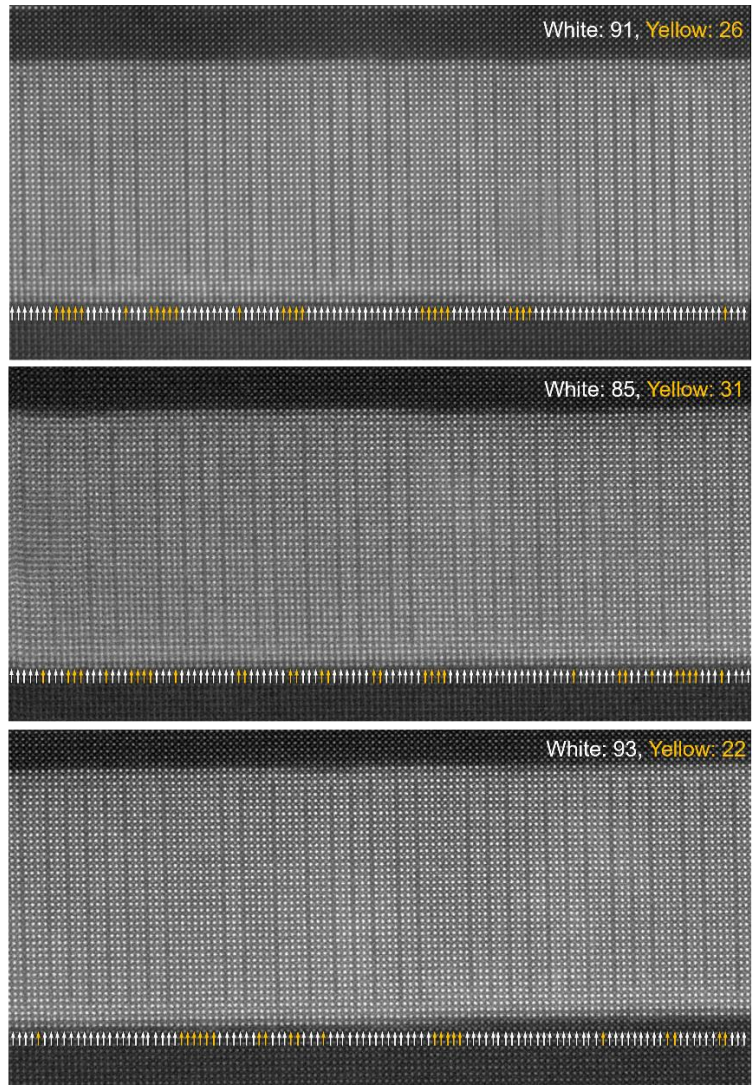

**Figure S5.** Volume fraction of twin-wall superlattice and bulk-like phase in an epitaxially strained LCO grown on a STO substrate. The white and yellow arrows indicate the structural units for the twin-wall superlattice and bulk-like phase, respectively. The number of total arrows are 348, while the number of the white and yellow arrows are 269 and 79. Accordingly, based on this HAADF STEM image analysis, 78% and 22% of the film interior correspond to the twin-wall superlattice and bulk-like phase, respectively.

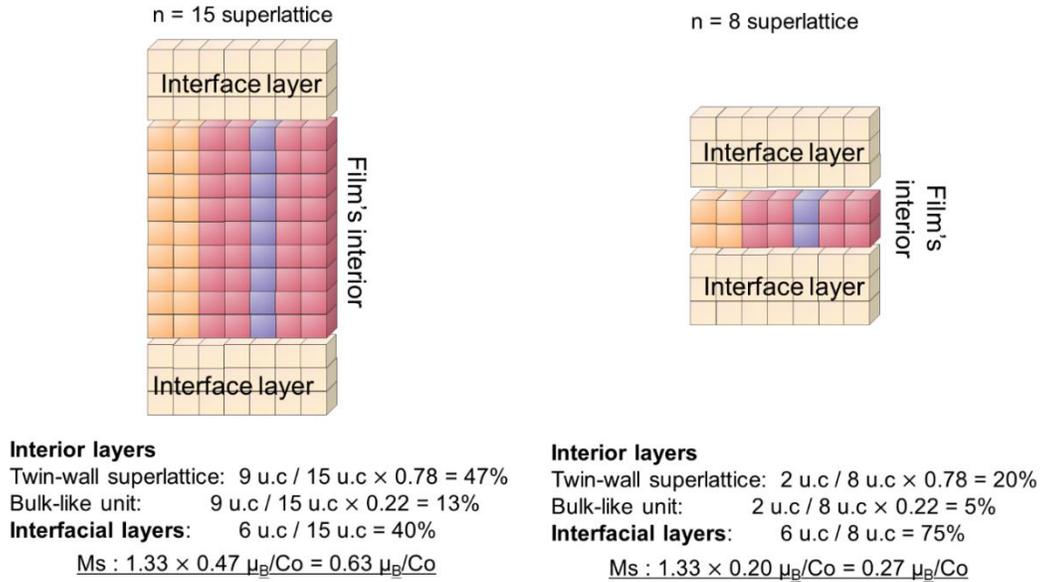

**Figure S6.** Volume fraction of twin-wall superlattice, bulk-like phase, and interfacial layers in n = 15 and 8 superlattices. The interfacial transition layers are shown to be 6 u.c. for each LCO layer sandwiched by STO layers, indicating that the remained layers in each LCO corresponds to the film's interior. To be specific, the thickness of the interior layers in $n = 15$ and 8 superlattices are 9 and 2 u.c., respectively. As demonstrated in Figure S5, the twin-wall superlattice and bulk-like phase occupied 78% and 22% of the film's interior. Therefore, the volume fraction of magnetically active twin-wall domain in $n = 15$ and 8 superlattices is 47% and 20%. HS/LS-FM model, the most stable magnetic configuration in DFT+U calculations, has the magnetic moment of 1.33 $\mu_B$/Co, suggesting that the saturated magnetic moments of $n = 15$ and 8 superlattices will be 0.63 and 0.27 $\mu_B$/Co. These values are consistent with the experimental $M_s$ measured by SQUID.

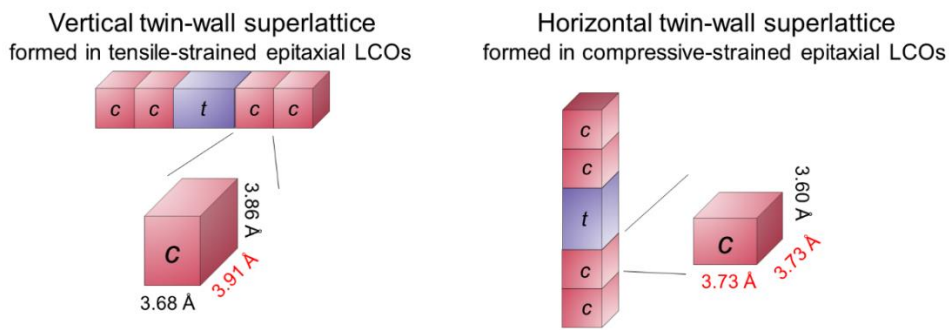

**Figure S7.** Lattice parameters of the *c*-units within vertical and horizontal twin-wall superlattice and magnetic easy axis of epitaxial strained LCOs. The lattice parameters of the c-unit measured by HAADF STEM are summarized above. The lattice parameters for horizontal twin-wall superlattice were found from the report from Ningbin *et al.( ACS applied materials & interfaces* **10**, 38230-38238 (2018)). The lattice parameter in depth direction is assumed to maintain the same lattice parameter of a substrate. The elongated axis of the c-units (denoted lattice sizes in red) is always in the in-plane direction, because the *c*-units are highly compressed in the propagation direction of the twin-wall superlattices. This is probably the reason that the easy-axis is always in the plane regardless of the strain direction or the orientation of the dark stripes.